\documentclass[journal]{IEEEtran}
\usepackage{amsmath,amsfonts}
\usepackage{algorithm}
\usepackage{array}
\usepackage[caption=false,font=normalsize,labelfont=sf,textfont=sf]{subfig}
\usepackage{textcomp}
\usepackage{stfloats}
\usepackage{url}
\usepackage{verbatim}
\usepackage{graphicx}

\usepackage{kotex}
\usepackage[T1]{fontenc}
\usepackage[utf8]{inputenc} 
\usepackage{xcolor}
\usepackage{algpseudocode}
\usepackage{booktabs}
\usepackage{amssymb}

\newtheorem{remark}{Remark}

\newcommand{\argmin}{{\arg\min}}

\newcommand{\rev}[1]{\textcolor{black}{#1}}
\newcommand{\edit}[1]{\textcolor{black}{#1}}

\begin{document}

\title{Code-Weight Sphere Decoding}

\author{Yubeen Jo, Geon Choi,~\IEEEmembership{Student Member,~IEEE}, Yongjune Kim,~\IEEEmembership{Member,~IEEE}, \\ and Namyoon Lee,~\IEEEmembership{Senior Member,~IEEE}
        
\thanks{Yubeen Jo is with the School of Electrical Engineering, Korea University, Seoul 02841, Republic of Korea (e-mail: jybin00@korea.ac.kr).}
\thanks{Geon Choi, Yongjune Kim and Namyoon Lee are with the Department of Electrical Engineering, POSTECH, Pohang 37673, Gyeongbuk, Republic of Korea (e-mail: geon.choi@postech.ac.kr; yongjune@postech.ac.kr; nylee@postech.ac.kr).}
}

\maketitle

\begin{abstract}
Ultra-reliable low-latency communications (URLLC) demand high-performance error-correcting codes and decoders in the finite blocklength regime. This letter introduces a novel two-stage near-maximum likelihood (near-ML) decoding framework applicable to any linear block code. Our approach first employs a low-complexity initial decoder. If this initial stage fails a cyclic redundancy check, it triggers a second stage: the proposed code-weight sphere decoding (WSD). WSD iteratively refines the codeword estimate by exploring a localized sphere of candidates constructed from pre-computed low-weight codewords. This strategy adaptively minimizes computational overhead at high signal-to-noise ratios while achieving near-ML performance, especially for low-rate codes. Extensive simulations demonstrate that our two-stage decoder provides an excellent trade-off between decoding reliability and complexity, establishing it as a promising solution for next-generation URLLC systems.
\end{abstract}


\section{Introduction}

\IEEEPARstart{U}{ltra-reliable} 
    low-latency communications (URLLC) necessitate efficient error-correcting codes at finite blocklengths and low code rates, as well as developing decoding algorithms that offer both low complexity and high performance \cite{shtblk_len_ch_code, effi_dec_sht_blk_len_6G, boss2, deep_polar_code,sparse_pretran_polar}. While significant research has focused on various suboptimal decoding methods such as syndrome decoding \cite{ecc_book}, ordered statistics decoding (OSD) \cite{osd}, guessing random additive noise decoding (GRAND) \cite{orbgrand, product-like}, and list decoding techniques such as successive cancellation list (SCL) decoding \cite{scl} including linearity-enhanced serial list decoding approaches that use sphere-based concepts \cite{linearity-enhanced}, achieving near-maximum likelihood (near-ML) decoding performance with optimized complexity at low code rates remains an open challenge.
    
    In this letter, we propose a novel two-stage near-ML decoding method specifically tailored for short blocklength and low-rate codes. The decoding process begins with an initial stage where a preliminary message estimate \rev{$\hat{\mathbf{m}}^{(-1)}$} is obtained using an arbitrary low-complexity suboptimal decoding algorithm. 
    
    \rev{If a cyclic redundancy check (CRC) is available, it serves as an efficient failure detector to determine the activation of the second stage. If the initial codeword satisfies the CRC validation, the process terminates immediately to minimize latency. However, it is important to note that the proposed method is applicable even without CRC. If the CRC check fails or is unavailable, the algorithm proceeds to the second stage: our introduced code-weight sphere decoding (WSD).}
    
    \rev{The core concept of WSD lies in initiating the search from the discrete re-encoded codeword estimate, rather than the continuous received signal used in conventional sphere decoding (SD) \cite{sphere}. By constructing a Hamming sphere $\mathcal{S}$ centered on the initial estimate $\hat{\mathbf{c}}^{(0)}$, WSD exploits the linearity of the code, where any neighbor can be represented as the sum of the initial codeword and a pre-computed low-weight error pattern.}
    
    \rev{Within this local sphere, WSD evaluates the Euclidean distance between the candidates and the received signal $\mathbf{y}$. Unlike the static tree search of SD, WSD employs an \textit{iterative hopping} strategy: if a candidate closer to $\mathbf{y}$ is found, the search center shifts to this new codeword, and the local search repeats, resembling a discrete gradient descent on the codeword lattice.}
    
    \rev{Furthermore, WSD incorporates a correlation-based filtering technique to significantly reduce computational complexity.} \rev{Even in the absence of CRC, if the initial decoder identifies the correct codeword, the WSD stage confirms the optimality of the current center and terminates within a single iteration, incurring negligible computational overhead.}
    
    A distinguishing feature of the proposed WSD decoder is its universal applicability to any linear code and any initial decoding method. Simulation results demonstrate that it achieves near-ML decoding performance across various low-rate codes in the short blocklength regime.

\section{Preliminaries}

    In this section, we define the fundamental notation. Let $|\mathcal{A}|$ denote the cardinality of a set $\mathcal{A}$, $\Vert\mathbf{v}\Vert$ the $L_2$ norm, and $\text{supp}({\bf c})$ the support of vector $\mathbf{c}$. We use $\mathbb{F}_2$ for the binary field and ${\bf 1}_N$ for the all-one vector. The Hamming distance and weight are denoted by $d_H(\cdot, \cdot)$ and $w_H(\cdot)$, respectively.

\subsection{Channel Coding System}
    We consider a binary linear block code $\mathcal{C}(N, K)$ with codeword length $N$ and input message length $K$.
    The code is defined by its generator matrix ${\bf G}\in \mathbb{F}_2^{K \times N}$. 
    An input vector ${\bf m} \in \mathbb{F}_2^{K}$ is mapped to a codeword ${\bf c} \in \mathbb{F}_2^N$ via the encoding relation ${\bf c} = {\bf m} {\bf G}$.

    \edit{When CRC precoding is employed to enhance error detection, the input dimension is set to $K_{\sf in} = K + K_{\sf crc}$, where $K_{\sf crc}$ is the number of CRC parity bits. In this case, the message vector ${\bf m} \in \mathbb{F}_2^K$ is first precoded into ${\bf v} = {\bf m} {\bf G}_{\sf crc} \in \mathbb{F}_2^{K_{\sf in}}$, and the channel encoding is performed with ${\bf m} = {\bf v}$.}
    
    For modulation, binary phase shift keying (BPSK) maps the binary codeword ${\bf c}$ to a symbol vector ${\mathbf x}(\mathbf{c}) = {\bf 1}_N-2{\bf c}\in \mathbb{R}^N$. This vector is transmitted over an additive white Gaussian noise (AWGN) channel, such that the received signal is ${\bf y} = {\bf x}(\mathbf{c}) + {\bf n}$, where ${\bf n} \sim \mathcal{N}(\mathbf{0}, \sigma^2\mathbf{I}_N)$ is an independent and identically distributed Gaussian noise vector with zero-mean and variance $\sigma^2=N_0/2$.
    
    Finally, the channel decoder reconstructs the estimated message $\hat {\mathbf{m}}$ from the received signal $\mathbf{y}$, aiming to minimize the decoding error probability.

\subsection{Code-Weight Sphere Construction}
    In a linear block code $\mathcal{C}$, the decoding search space can be efficiently structured by decomposing the code lattice into \edit{\textit{Hamming shells} based on the code weights. 
    For any reference codeword ${\bf c} \in \mathcal{C}$, we define the Hamming shell of weight $d_\ell$ as $\mathcal{C}_{d_\ell} ({\bf c}) = \{ {\bf c}' \in \mathcal{C}: d_H({\bf c}, {\bf c}') = d_\ell\}$, where $d_\ell$ represents the $\ell$-th distinct value in the code's weight spectrum.}
    Suppose the support of the weight spectrum consists of $L+1$ distinct elements arranged in increasing order, i.e., $d_0 (= 0) < d_1 (= d_{\min}) < \dots < d_L$, where $d_{\min}$ denotes the minimum Hamming distance. The entire codeword set $\mathcal{C}$ can be decomposed into disjoint Hamming shells.
    
\begin{figure}[t]
      \centering
      \includegraphics[width=\columnwidth]{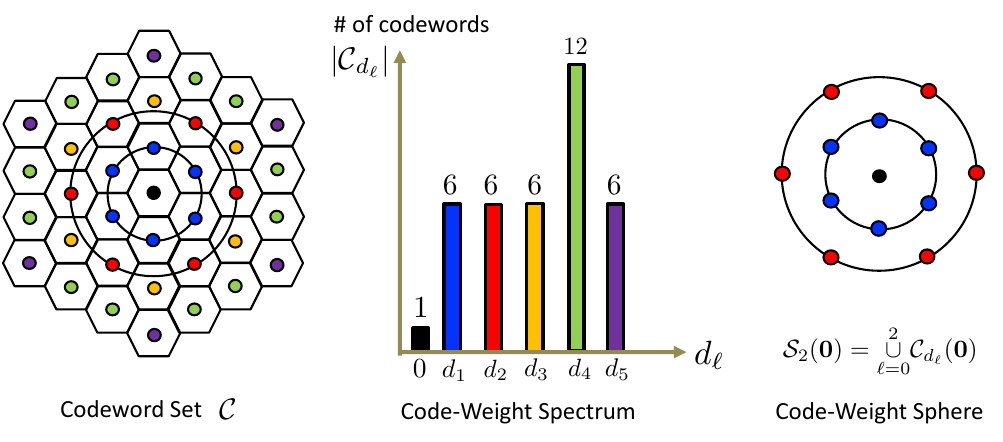}
      \caption{\edit{Illustration of the codeword set $\mathcal{C}$, code-weight spectrum, and code-weight sphere construction.}}
      \label{fig:Hamming_sphere}
      \vspace{-3mm}
    \end{figure}

    Building on this definition, we define the \textit{code-weight sphere} centered at ${\bf c}$ with a radius index $r$ (where $d_r \le d_L$), denoted as $\mathcal{S}_r({\bf c})$, as the \edit{union of Hamming shells with distances ranging from $d_0$ to $d_r$:
    \begin{align}
    	\mathcal{S}_r({\bf c}) = \bigcup_{\ell=0}^{r} \mathcal{C}_{d_\ell}({\bf c}).
    \end{align}}
    Since $\mathcal{C}_{d_0}({\bf c}) = \{{\bf c}\}$, the sphere includes the center codeword itself.
    Note that $\mathcal{S}_L({\bf c})$ is equivalent to the codebook $\mathcal{C}$. Fig. \ref{fig:Hamming_sphere} illustrates this construction.

    A fundamental property of linear codes is that the local geometry around any codeword is identical to the geometry around the zero codeword ${\bf 0}$.
    Specifically, the Hamming shell $\mathcal{C}_{d_\ell}(\hat {\bf c})$ around an arbitrary estimate $\hat {\bf c}$ forms a coset of the shell around the zero codeword $\mathcal{C}_{d_\ell}({\bf 0})$.
    Since $d_H(\hat{\bf c}, {\bf c}) = d_H({\bf 0}, {\bf c}-\hat{\bf c}) = w_H({\bf c}-\hat{\bf c})$ (in binary fields, subtraction is equivalent to addition), the following relationship holds:
    \begin{align}
    	\mathcal{C}_{d_\ell}(\hat {\bf c}) 
    	&= \{ \hat{\bf c} + {\bf c}: {\bf c} \in \mathcal{C}_{d_\ell} ({\bf 0}) \} \nonumber \\
    	&= \hat{\bf c} + \mathcal{C}_{d_\ell}({\bf 0}),
    \end{align}
    where $\mathcal{C}_{d_\ell}({\bf 0})$ represents the set of all codewords with a Hamming weight of exactly $d_\ell$.
    Consequently, the code-weight sphere around $\hat{\mathbf{c}}$ is simply the translation of the sphere around the zero codeword: $\mathcal{S}_r(\hat{\mathbf{c}}) = \hat{\mathbf{c}} + \mathcal{S}_r(\mathbf{0})$.
    This affine property significantly reduces computational complexity, as it allows the decoder to pre-compute and store only the low-weight codewords in $\mathcal{S}_r({\bf 0})$, rather than constructing spheres dynamically for every estimate.
    
    \section{A Two-Stage Near-ML Decoder}
    In this section, we present a two-stage decoding method that leverages both a low-complexity decoder and WSD.

\subsection{Initial Low-Complexity Decoding Phase}  
    In the initial phase, a low-complexity decoder generates a preliminary codeword estimate $\hat{\mathbf{c}}^{(-1)}$. 
    If a CRC is employed, the decoder first validates the extracted vector $\hat {\bf v}^{(-1)}$; the process terminates early if the CRC validation passes (i.e., ${{\bf H}_{\sf crc} (\hat{\bf  v}^{(-1)})^\top={\bf 0}}$), significantly reducing the average computational complexity.
    
    \rev{In the event of a CRC failure or when a CRC is not available, the system proceeds to the WSD phase. 
    First, the message bit vector ${\hat {\bf m}}^{(-1)}$ is extracted from the initial estimate (either from $\hat{\mathbf{c}}^{(-1)}$ or $\hat {\bf v}^{(-1)}$) and is re-encoded to form the reference codeword $\hat{\mathbf{c}}^{(0)}$. 
    Specifically, $\hat{\mathbf{c}}^{(0)} = {\hat{\bf  m}^{(-1)}}{\bf G}$ for non-CRC cases, while $\hat{\mathbf{c}}^{(0)} = ({\hat{\bf  m}^{(-1)}}{\bf G}_{\sf crc}){\bf G}$ for CRC-aided cases. 
    Finally, we define the reliability metric at the $i$-th iteration as $M^{(i)} \triangleq \|{\bf y} - {\bf x}(\hat{\mathbf{c}}^{(i)})\|$. The process begins by computing the initial reliability $M^{(0)}$ from $\hat{\mathbf{c}}^{(0)}$, which serves as the baseline for the subsequent boosting iterations.}

\subsection{Code-Weight Sphere Decoding Phase}
\begin{figure}[t]
  \centering
  \includegraphics[width=\columnwidth]{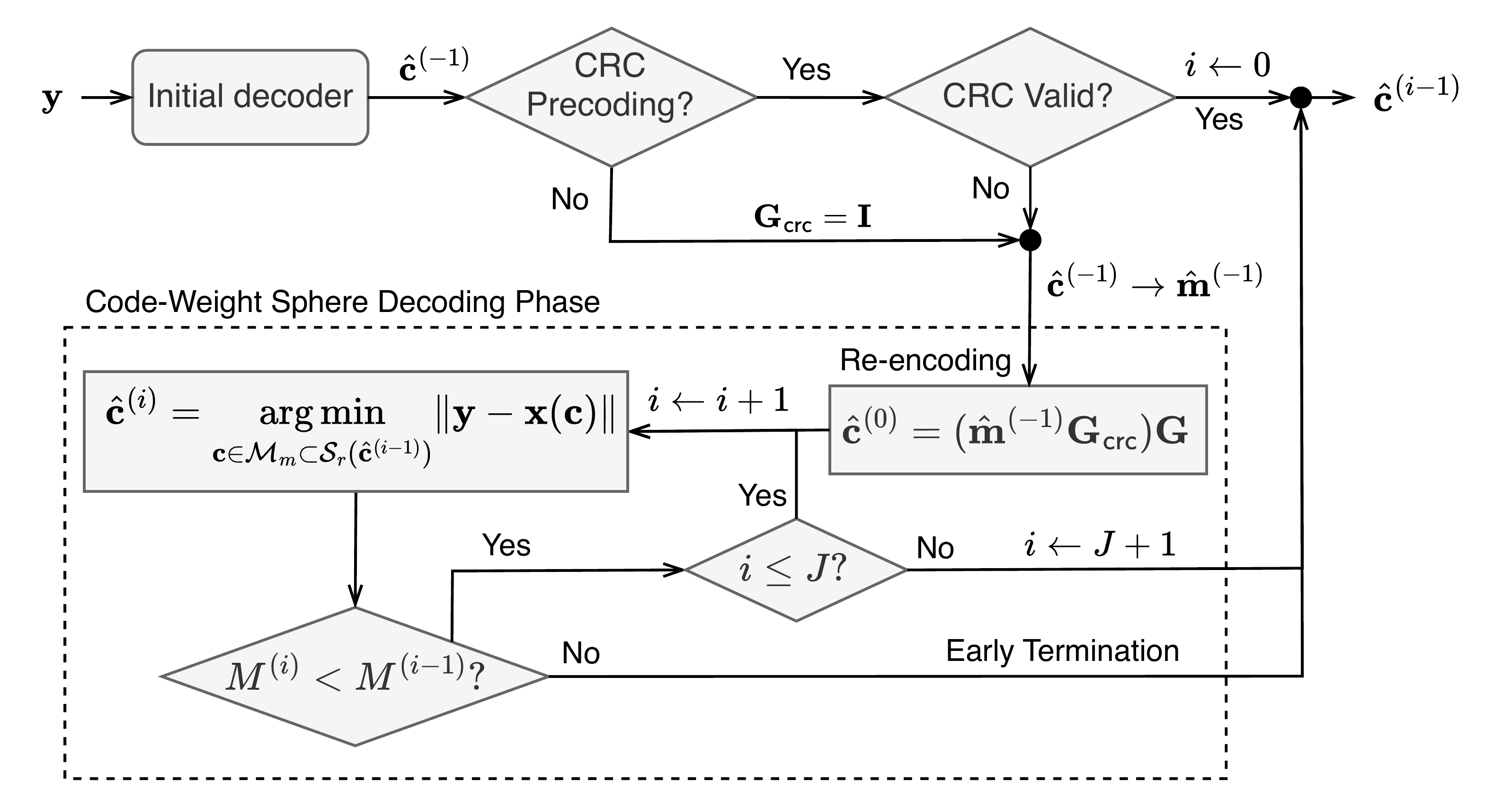}
  \caption{\rev{Flowchart of the proposed two-stage decoding algorithm, detailing the code-weight sphere decoding (WSD) phase.}}
  \label{fig:WSD_alg}
  \vspace{-3mm}
\end{figure}

    The WSD phase encompasses multiple iterations designed to refine the initial codeword estimate $\hat{\bf c}^{(0)}$. 
    During this phase, the decoder iteratively searches for a better candidate within the code-weight sphere set $\mathcal{S}_{r}(\hat{\bf c}^{(i-1)})$, up to a maximum of $J$ iterations.
    In the $i$-th boosting round ($i \ge 1$), the decoder identifies a candidate ${\bf c}^*$ that minimizes the Euclidean distance to the received signal ${\bf y}$:
    \begin{align}
        {\bf c}^* = \underset{{{\bf c} \in \mathcal{S}_{r}(\hat{\mathbf{c}}^{(i-1)})}}{\argmin} \|{\bf y} - {\bf x}({\bf c})\|.
        \label{eq:wsd_optimization}
    \end{align}
    \rev{In the practical implementation shown in Fig. \ref{fig:WSD_alg}, the minimization in \eqref{eq:wsd_optimization} is performed over a filtered subset $\mathcal{M}_m \subset \mathcal{S}_r(\hat{\mathbf{c}}^{(i-1)})$, which consists of the top-$m$ candidates identified by the correlation metric to reduce complexity.}
    
    \rev{The decoder then evaluates the reliability of this new candidate. If the Euclidean distance metric improves (i.e., $\|{\bf y} - {\bf x}({\bf c}^*)\| < M^{(i-1)}$), the estimate is updated to $\hat{\mathbf{c}}^{(i)} = {\bf c}^*$ and the reliability metric becomes $M^{(i)} = \|{\bf y} - {\bf x}({\bf c}^*)\|$. Then, the process advances to the next iteration. Otherwise, the decoding terminates, returning $\hat{\mathbf{c}}^{(i-1)}$ as the final decision. Fig. \ref{fig:WSD_alg} provides a comprehensive flowchart of this procedure.}

\begin{remark}[Decoding Stability and Refinement]
	By construction, the proposed decoder ensures a strictly monotonic improvement in reliability (i.e., $M^{(0)} > M^{(1)} > \dots > M^{(J)}$). 
	\rev{This greedy update prevents error accumulation and divergence. Consequently, the decoder either converges to a local optimum or terminates early, guaranteeing that performance remains at least equal to the initial decoding stage.}
\end{remark}

\begin{remark}[Difference from Sphere Decoding]
	\rev{While both SD and WSD pursue the ML solution, they differ fundamentally. First, unlike SD which centers the search on the continuous signal $\mathbf{y}$ (requiring online tree construction), WSD initiates from a discrete codeword estimate. This allows WSD to exploit code linearity and construct the search space instantly using \textit{offline pre-computed} spheres. Second, instead of a static tree traversal, WSD employs a dynamic \textit{iterative hopping} strategy (similar to discrete gradient descent). Finally, WSD enforces a strict latency bound via the iteration parameter $J$, whereas SD's complexity varies with channel noise.}
\end{remark}

\subsection{Decoding Complexity}
\rev{To ensure a fair comparison across different decoding architectures, we evaluate the computational complexity in terms of floating-point operations (FLOPs) and normalize them into equivalent \textit{Euclidean distance} (ED) \textit{units}.
We define one ED unit as the cost of computing $\|\mathbf{y} - \mathbf{x}\|^2$ for a vector of length $N$, which requires approximately $3N$ FLOPs.
Consequently, the complexity of the brute-force ML decoder is normalized to $2^K$ ED units. }

\rev{For the benchmark SCL decoder, we model the complexity of log-likelihood ratio (LLR)-based SCL decoding as $\mathcal{C}_{\sf scl} \approx \frac{4}{3} L \log_2 N$, assuming 4 FLOPs per processing node. 
This is extended to SCL-back propagation parity check (SCL-BPC) \cite{deep_polar_code} by adding the pre-transformation overhead, modeled as $\mathcal{C}_{\sf bpc} \approx \mathcal{C}_{\sf scl} + \frac{4}{3N} \sum_\ell N_\ell \log_2 N_\ell$. 
The OSD complexity is approximated by the order-$k$ reprocessing stage, $\mathcal{C}_{\sf osd} \approx \sum_{i=0}^k \binom{K}{i}$. Detailed formulations are summarized in Table \ref{tab:complexity}.}

\rev{In the proposed WSD phase, calculating the exact ED for every candidate in $\mathcal{S}_r(\mathbf{0})$ is inefficient. To address this, we employ a \textit{correlation-based filtering} technique. Since minimizing the ED $\|\mathbf{y} - \mathbf{x}\|^2 = \|\mathbf{y}\|^2 + \|\mathbf{x}\|^2 - 2\mathbf{y}^\top\mathbf{x}$ is equivalent to maximizing the correlation $\mathbf{y}^\top\mathbf{x}$ for constant-energy BPSK signals, we utilize a simplified gain metric $\mathcal{G}(\mathbf{c})$. This metric represents the incremental improvement (gain) in the correlation when shifting the current search center $\hat{\mathbf{x}}^{(i-1)}$ by a low-weight codeword $\mathbf{c}$, derived as a sparse summation: $\mathcal{G}(\mathbf{c}) = \sum_{j \in \text{supp}(\mathbf{c})} -2 y_j \hat{x}^{(i-1)}_j$.}

\rev{By identifying candidates with the highest correlation gain $\mathcal{G}(\mathbf{c})$, WSD effectively screens for codewords most likely to be closer to $\mathbf{y}$ without full distance calculations. This operation requires only $w_H(\mathbf{c})$ additions, which is significantly cheaper than the $3N$ FLOPs required for an exact ED.}

\rev{Based on this metric, WSD selects the top-$m$ candidates (forming the subset $\mathcal{M}_m$) for exact ED verification.
Considering the filtering and sorting overheads, the worst-case normalized complexity of WSD is bounded by:
\begin{align}
	\mathcal{C}_{\sf wsd} \approx J \times \left(m\left(1+\frac{1}{3N}\right)+ \frac{|\mathcal{S}_r(\mathbf{0})|(\bar{w} + \log_2 m)}{3N} \right),
    \label{eq:wsd_comp}
\end{align}
where $\bar{w}$ denotes the average Hamming weight of the candidate codewords in $\mathcal{S}_r(\mathbf{0})$ and $m$ is the filtering threshold size.}
\rev{Finally, the total average complexity is given by:
\begin{align}
	\mathcal{C}_{\sf avg} = \mathcal{C}_{\sf init} + P_{\sf act} \times \mathcal{C}_{\sf wsd},
\end{align}
where $P_{\sf act}$ denotes the activation probability of the WSD phase. For CRC-aided systems, $P_{\sf act} = P_{\sf e,crc}$ represents the probability of a CRC failure. For scenarios without CRC (e.g., the Reed-Muller (RM) code with OSD), $P_{\sf act}$ is effectively set to $1$, as the WSD stage can be configured to operate in an \textit{always-on mode} (AOM) to guarantee reliability refinement.}

\rev{At high singal-to-noise ratio (SNR), $P_{\sf e,crc} \approx 0$, making the overhead negligible. At low SNR, practical feasibility is ensured via three mechanisms:
\begin{itemize}
    \item \textbf{Bounded Latency}: The worst-case delay is deterministically limited by $J$.
    \item \textbf{Parallelizability}: Operations within the filtering and ED stages are mutually independent across candidates, allowing massive hardware parallelism.
    \item \textbf{Adaptivity}: WSD can be selectively disabled if the estimated SNR falls below a threshold to prevent unnecessary overhead.
\end{itemize}}

\begin{table}[htbp]
\renewcommand{\arraystretch}{1.2}
\centering
\caption{Theoretical Normalized Complexity Analysis (in ED units)}
\label{tab:complexity}
\footnotesize
\begin{tabular}{@{}ll@{}}
\toprule
Decoder & \multicolumn{1}{c}{Normalized Complexity (Approx.)} \\ 
\midrule
SCL$(L)$            & $\frac{4}{3} L \log_2 N$  \\ 
SCL$(L)$ + WSD$(r)$ & $\mathcal{C}_{\sf scl} + P_{\sf act}\times \mathcal{C}_{\sf wsd}$ \\ 
\midrule
SCL-BPC$(L)$        & $\mathcal{C}_{\sf scl} + \frac{4}{3N} \sum_{\ell} N_\ell \log_2 N_\ell$ \\
SCL-BPC$(L)$ + WSD$(r)$ & $\mathcal{C}_{\sf bpc} + P_{\sf act}\times \mathcal{C}_{\sf wsd}$ \\
\midrule
OSD$(k)$            & $\sum_{i=0}^k \binom{K}{i}$ \\
OSD$(k)$ + WSD$(r)$ & $\mathcal{C}_{\sf osd} + \mathcal{C}_{\sf wsd}$ \\
\midrule
MLD & $2^K$ \\ 
\bottomrule
\multicolumn{2}{@{}p{0.98\columnwidth}@{}}{
    \footnotesize * $P_{\sf act}$: Activation probability ($P_{\sf act} = P_{\sf e,crc}$ for CRC-aided cases; $P_{\sf act} = 1$ for non-CRC cases).
}
\end{tabular}
\vspace{-5mm}
\end{table}

\section{Simulation Results}\label{sec:simulation}

We evaluate the block error rate (BLER) and computational complexity of the proposed WSD under a binary-input AWGN channel.

\subsection{Simulation Settings and Benchmarks}

We compare WSD against various state-of-the-art decoding schemes. The benchmark settings are summarized as follows:
\begin{itemize}
    \item \textbf{CRC-aided (CA) polar codes:} We employ CA-polar codes constructed using the 5th generation new radio (5G-NR) reliability sequence \cite{3gpp2020}. CRC precoding is performed with the generator polynomial $g(x) = 1+x^5+x^9+x^{10}+x^{11}$. We use successive cancellation list (SCL) decoders with list sizes $L \in \{8, 32\}$.
    
    \item \textbf{CA-deep polar (CA-DP) codes:} For CA-DP codes \cite{deep_polar_code}, we utilize the SCL-BPC decoder ($L=32$). The 5G-NR sequence is adopted for information bit selection with CRC polynomial $g(x) = 1 + x^5 + x^6$.

    \item \rev{\textbf{OSD:} To verify universality, we evaluate the $\mathcal{RM}$ code $\mathcal{RM}(128, 29)$ using the OSD algorithm \cite{osd} with order $k \in \{2, 3, 4\}$.}
    
    \item \textbf{Proposed WSD:} WSD utilizes the pre-computed sphere set $\mathcal{S}_r(\mathbf{0})$. 
    \rev{Based on our analysis of the sphere coverage via Minkowski sums and empirical simulations, we observed that the performance gain saturates around $J=4$. To maximize the performance-to-complexity ratio, the maximum iterations is set to $J=4$.}
    \rev{Regarding the filtering parameter, we set $m=\max(100,\lceil0.02\times|\mathcal{S}_r(\mathbf{0})|\rceil)$. Empirical results confirm that this 2\% threshold incurs no performance degradation, and for small sets ($|\mathcal{S}_r(\mathbf{0})|$ < 100), the filtering is bypassed to ensure optimality.}
    
    \item \textbf{Reference:} The ML decoder (MLD) and theoretical bounds (random coding union (RCU) bound \cite{rcu}, Meta-converse bound \cite{meta_conv}) are provided as performance limits.
\end{itemize}
Simulation parameters are detailed in Table \ref{tab:simulation_parameters}, where the specific code-weight sphere sizes employed for each $(N, K)$ configuration are highlighted in bold.

\begin{table}[htbp]
    \footnotesize 
    \footnotesize 
    \centering
    \caption{Cardinality of the code-weight sphere $\mathcal{S}_r({\mathbf{0}})$ in Sec.~\ref{sec:simulation}}
    \label{tab:simulation_parameters}
    
    \setlength{\tabcolsep}{3.5pt} 
    
    \begin{tabular*}{\columnwidth}{@{\extracolsep{\fill}}lllcccc@{}}
        \toprule
        & $(N,K)$ & Fig. & $|\mathcal{S}_1|$ & $|\mathcal{S}_2|$ & $|\mathcal{S}_3|$ & $|\mathcal{S}_4|$ \\ 
        \midrule
        CA-polar &$(64,16)$ & \ref{fig:sim_scl_wsd} &$\bf 9$ & $\bf 246$ & $\bf 4,002$ & $-$\\ 
        CA-polar &$(128,16)$ &\ref{fig:sim_scl_wsd} & $1$ & $\bf 24$ & $\bf 1,078$ & $\bf 12,995$ \\ 
        CA-polar &$(256,16)$ &\ref{fig:sim_scl_wsd} & $1$ & $\bf 10$ & $\bf 537$ & $\bf 6,471$\\ 
        \midrule
        CA-DP  &$(64,16)$&  \ref{fig:sim_deep_polar_wsd}& $8 $& $277$ & $\bf 3,941$ & $-$ \\ 
        CA-DP  &$(128,16)$  & \ref{fig:sim_deep_polar_wsd}& $610$ & $\bf 14,490$ & $-$ & $-$ \\ 
        CA-DP  &$(256,16)$  & \ref{fig:sim_deep_polar_wsd}& $214$ & $5,670$ & $\bf 13,222$ & $-$\\ 
        \midrule
        $\mathcal{RM}$ & $(128,29)$ & \ref{fig:sim_osd_rm_wsd} & $\bf 10,688$ & $-$&$-$ &$-$ \\
        \bottomrule
    \end{tabular*}
    \vspace{-3mm} 
\end{table}

\begin{figure}[tbph]
  \centering
  \includegraphics[width=\columnwidth]{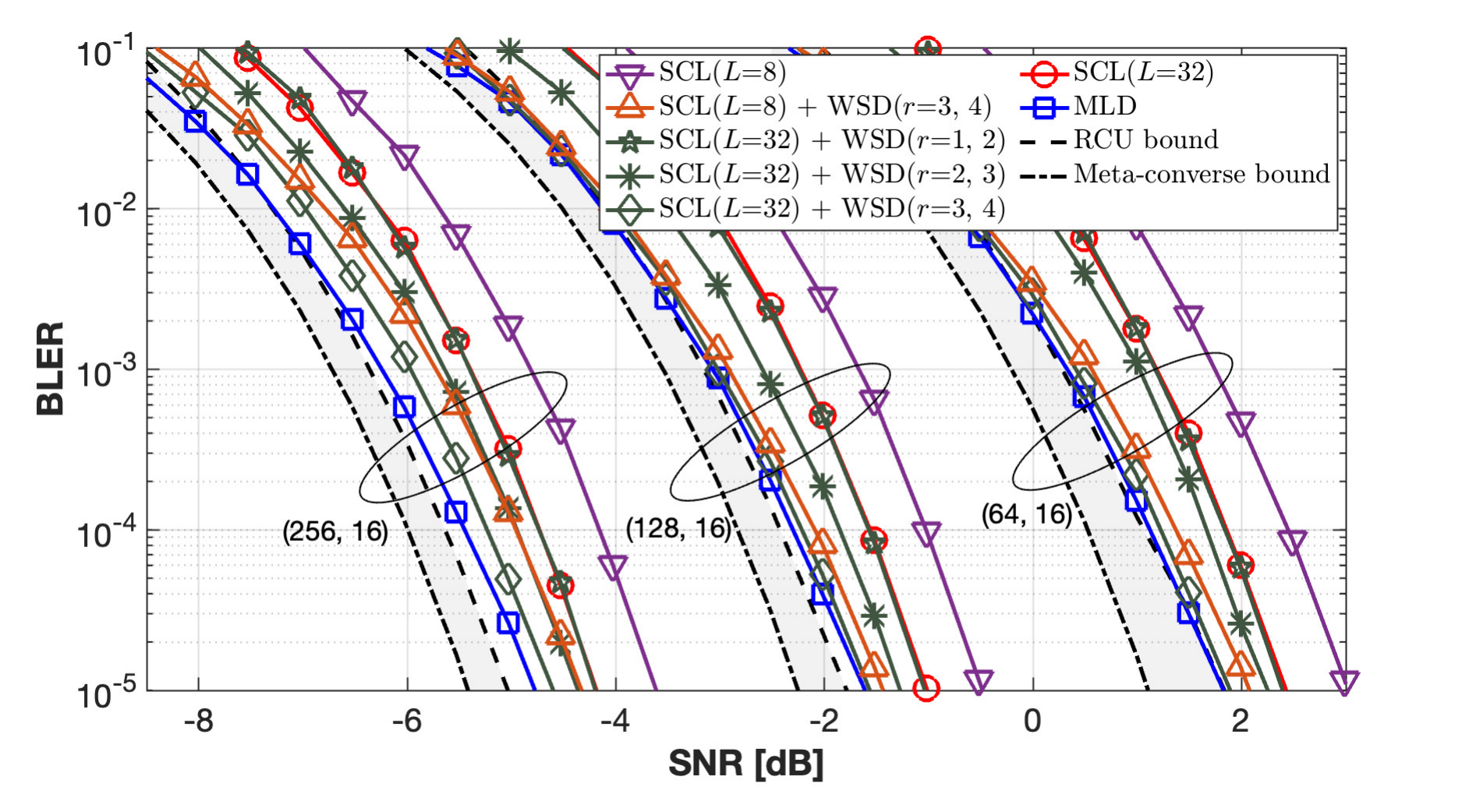}
  \vspace{-6mm}
  \caption{\rev{BLER performance comparison for CA-polar codes, demonstrating the impact of both list size ($L$=8, 32) and code-weight sphere size ($r$).}}
  \label{fig:sim_scl_wsd}
  \vspace{-5mm}
\end{figure}

\begin{figure}[tbph]
  \centering
  \includegraphics[width=\columnwidth]{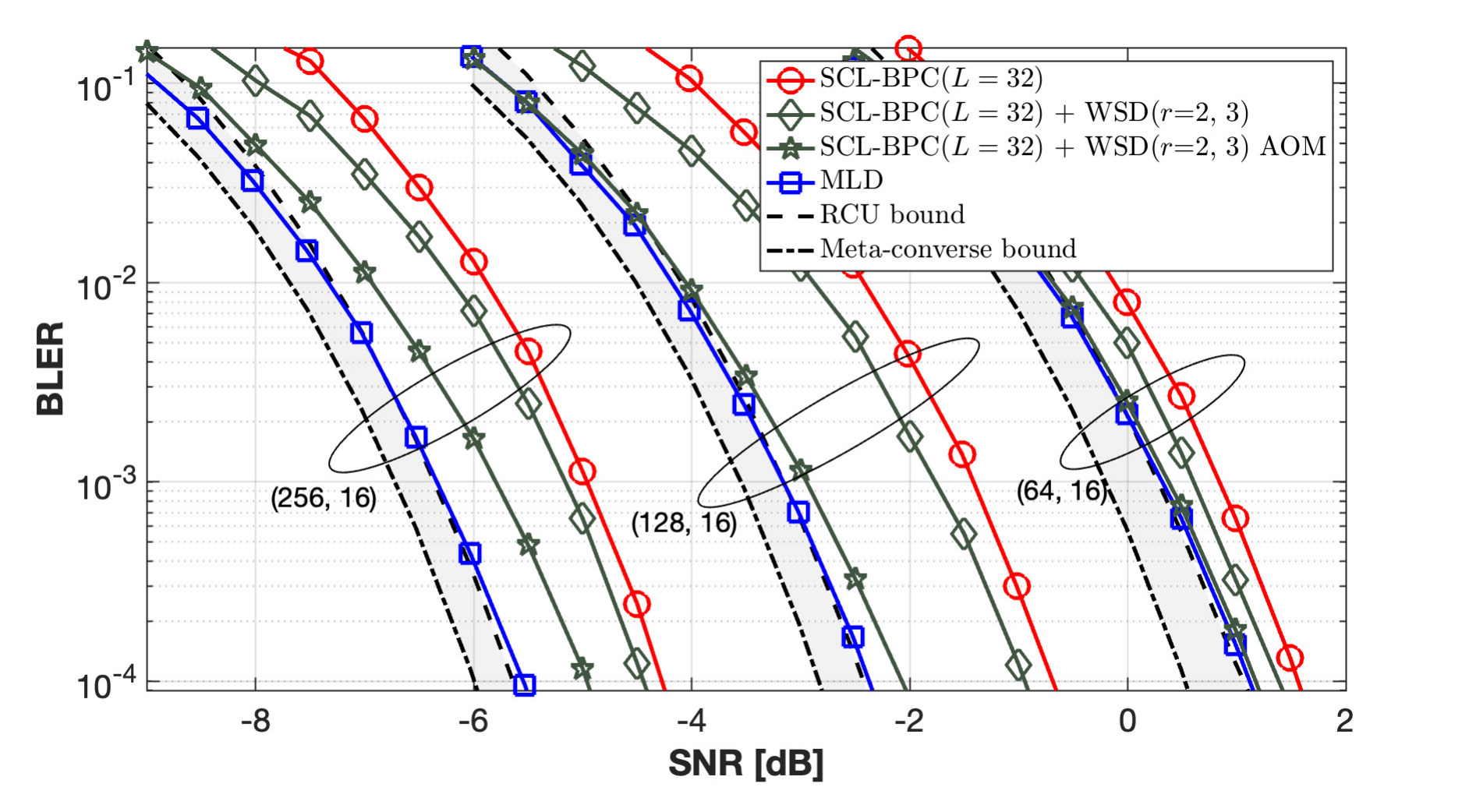}
  \vspace{-6mm}
  \caption{BLER comparison for CA-DP codes under SCL-BPC, SCL-BPC+WSD, and MLD decoding methods.}
  \label{fig:sim_deep_polar_wsd}
  \vspace{-3mm}
\end{figure}

\begin{figure}[tbph]
  \centering
  \includegraphics[width=\columnwidth]{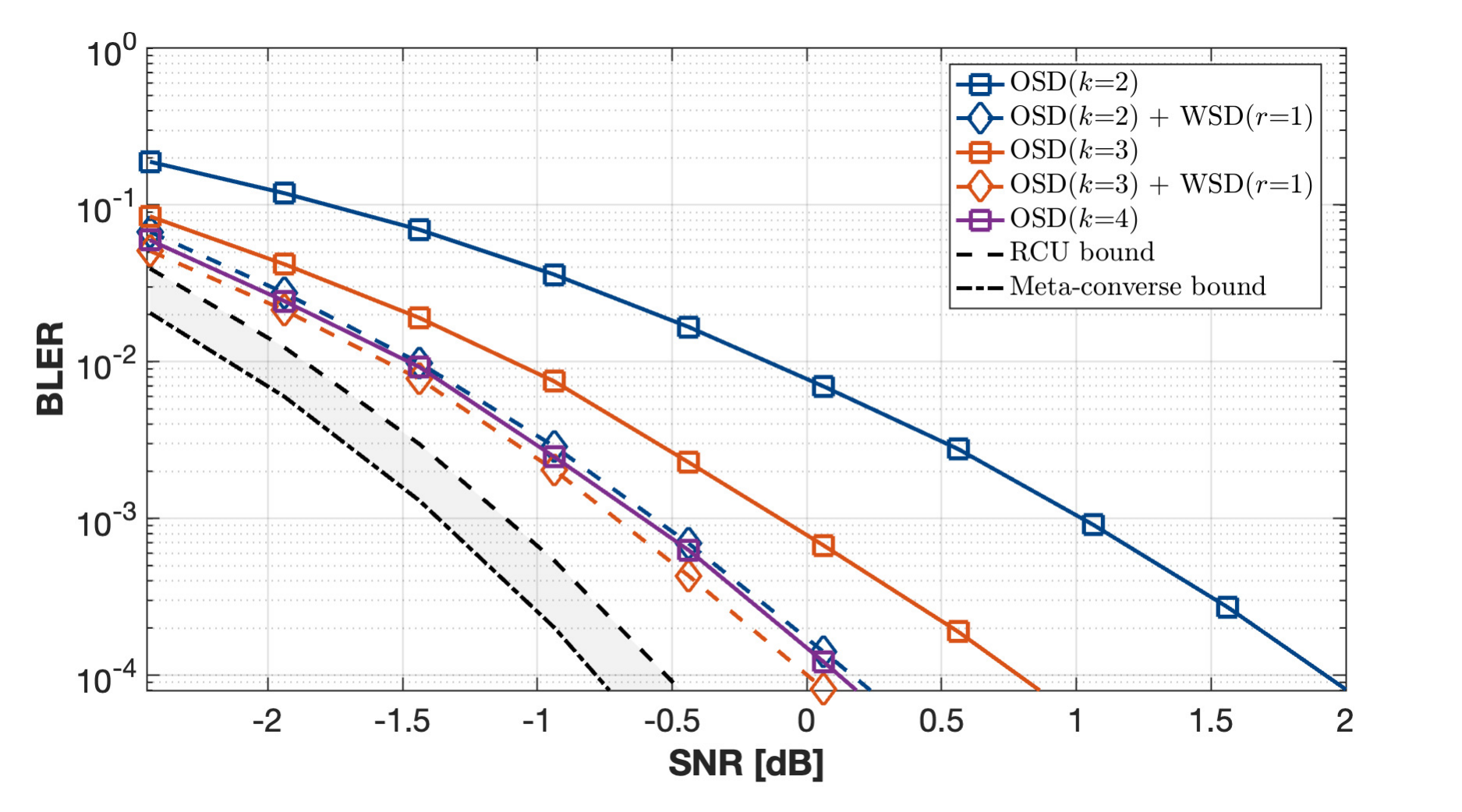}
  \vspace{-6mm}
  \caption{\rev{BLER comparison for the RM(128, 29) code under OSD and OSD+WSD decoding methods.}}
  \label{fig:sim_osd_rm_wsd}
  \vspace{-3mm}
\end{figure}

\subsection{Comparison of BLER Performance and Decoding Complexity}

We evaluate the performance for blocklengths $N \in \{64, 128, 256\}$ with code rate $R=\{\frac{1}{4}, \frac{1}{8}, \frac{1}{16}\}$.
Fig. \ref{fig:sim_scl_wsd} illustrates how the BLER performance improves as the search radius $r$ increases. For CA-polar codes with $L=32$, SCL+WSD with $r \ge 3$ becomes nearly indistinguishable from MLD for $N=64, 128$. Even for $N=256$, the gap to the ML bound is negligible ($<0.2$ dB at $10^{-5}$ BLER), demonstrating that the performance systematically converges to the ML limit by scaling the set size $r$.

\begin{figure}[htbp]
  \centering
  \includegraphics[width=\columnwidth]{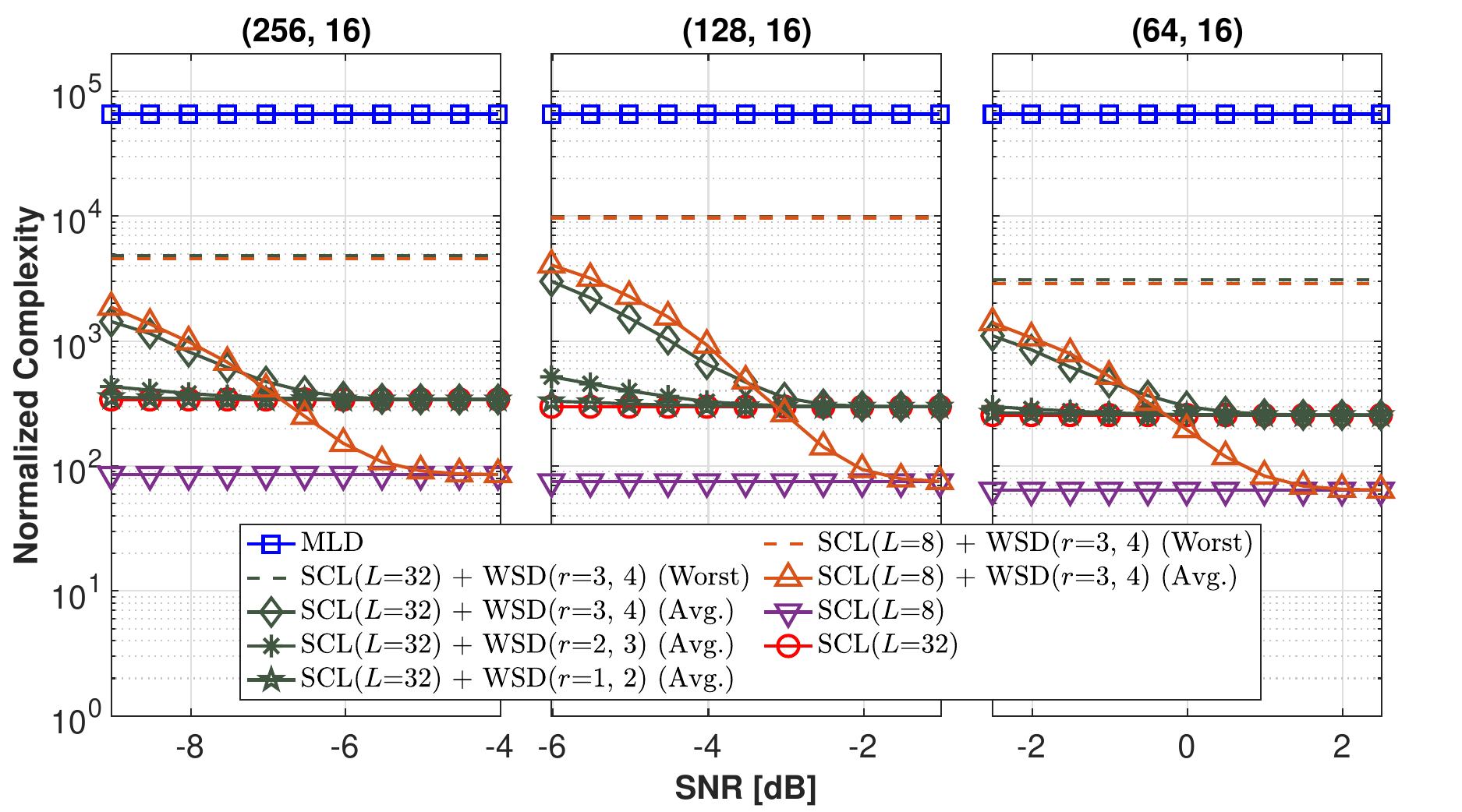}
  \caption{\rev{Average complexity comparison for CA-polar codes ($K=16$) relative to list size ($L$) and code-weight sphere size ($r$).}}
  \label{fig:sim_scl_wsd_comp}
  \vspace{-3mm}
\end{figure}

\rev{We specifically investigated the impact of the initial list size $L$ on decoding efficiency. Figs. \ref{fig:sim_scl_wsd} and \ref{fig:sim_scl_wsd_comp} reveal that SCL($L=8$)+WSD not only matches the BLER performance of the standalone SCL($L=32$) but also approaches the near-ML performance of SCL($L=32$)+WSD. Regarding complexity, while the overhead of SCL($L=8$)+WSD is comparable to that of SCL($L=32$)+WSD at low SNRs due to frequent WSD activation, it dynamically converges to the baseline complexity of SCL($L=8$) as the SNR increases. This confirms that WSD can effectively compensate for a smaller initial list size, providing a superior performance-complexity trade-off.}

Similarly, for CA-DP codes$-$a class of $\mathcal{RM}$-type pre-transformed polar codes (Fig. \ref{fig:sim_deep_polar_wsd})$-$WSD consistently enhances the state-of-the-art SCL-BPC decoder. The AOM ($P_{\sf act}=1$) results reveal a substantial performance gain exceeding 1.2 dB for $N=128$, confirming that WSD effectively refines estimates even beyond the error detection limits of the CRC.

\rev{To validate universality, we applied WSD to the $\mathcal{RM}$(128, 29) code. As shown in Fig. \ref{fig:sim_osd_rm_wsd}, WSD ($r=1$) consistently improves the BLER of OSD ($k=2, 3$), confirming its effectiveness across different code structures.}

\begin{figure}[htbp]
  \centering
  \includegraphics[width=\columnwidth]{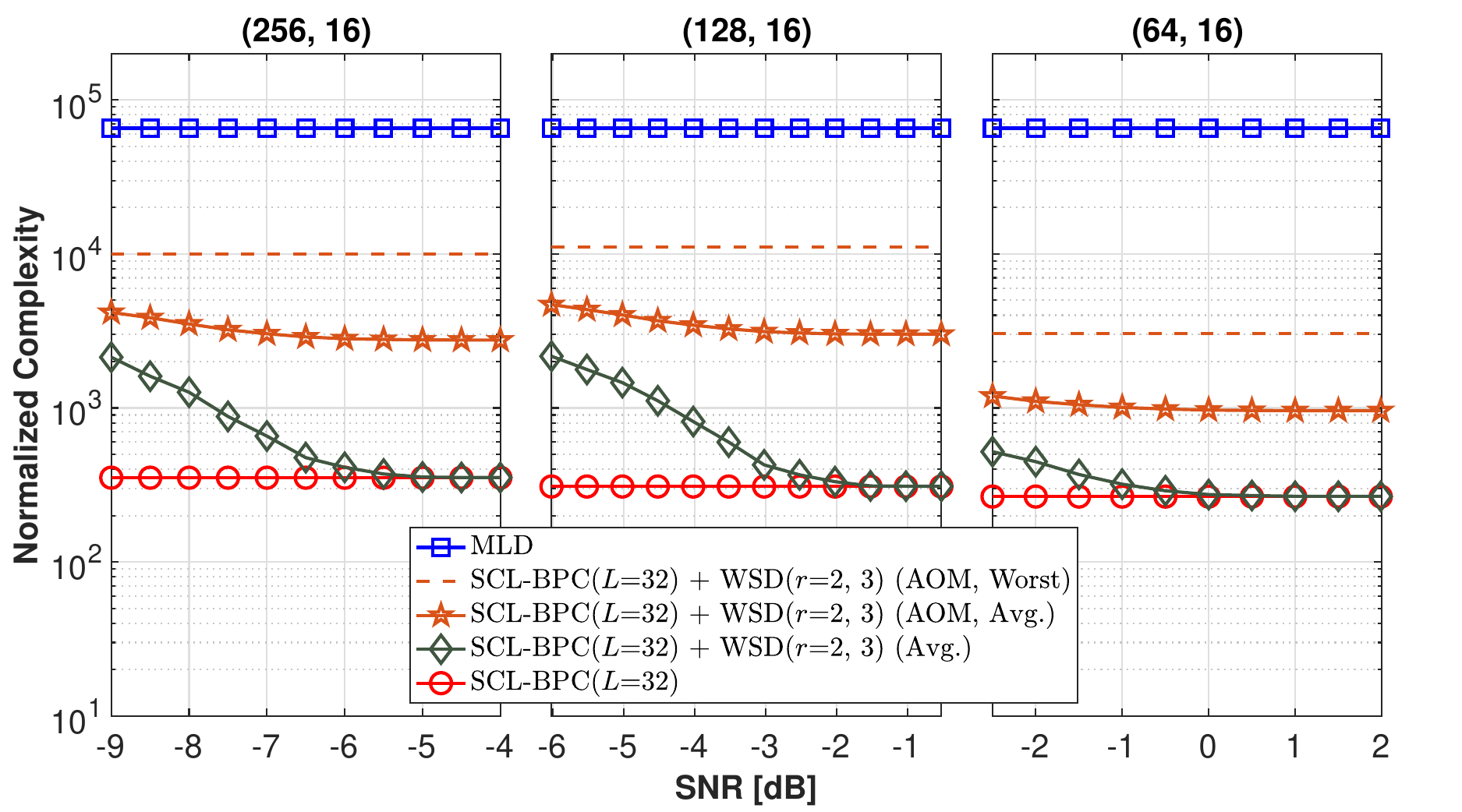}
  \caption{\rev{Average complexity comparison for CA-DP codes ($K=16$) under SCL-BPC, SCL-BPC+WSD, and MLD decoding methods.}}
  \label{fig:sim_scl_bpc_wsd_comp}
  \vspace{-3mm}
\end{figure}

\begin{figure}[htbp]
  \centering
  \includegraphics[width=\columnwidth]{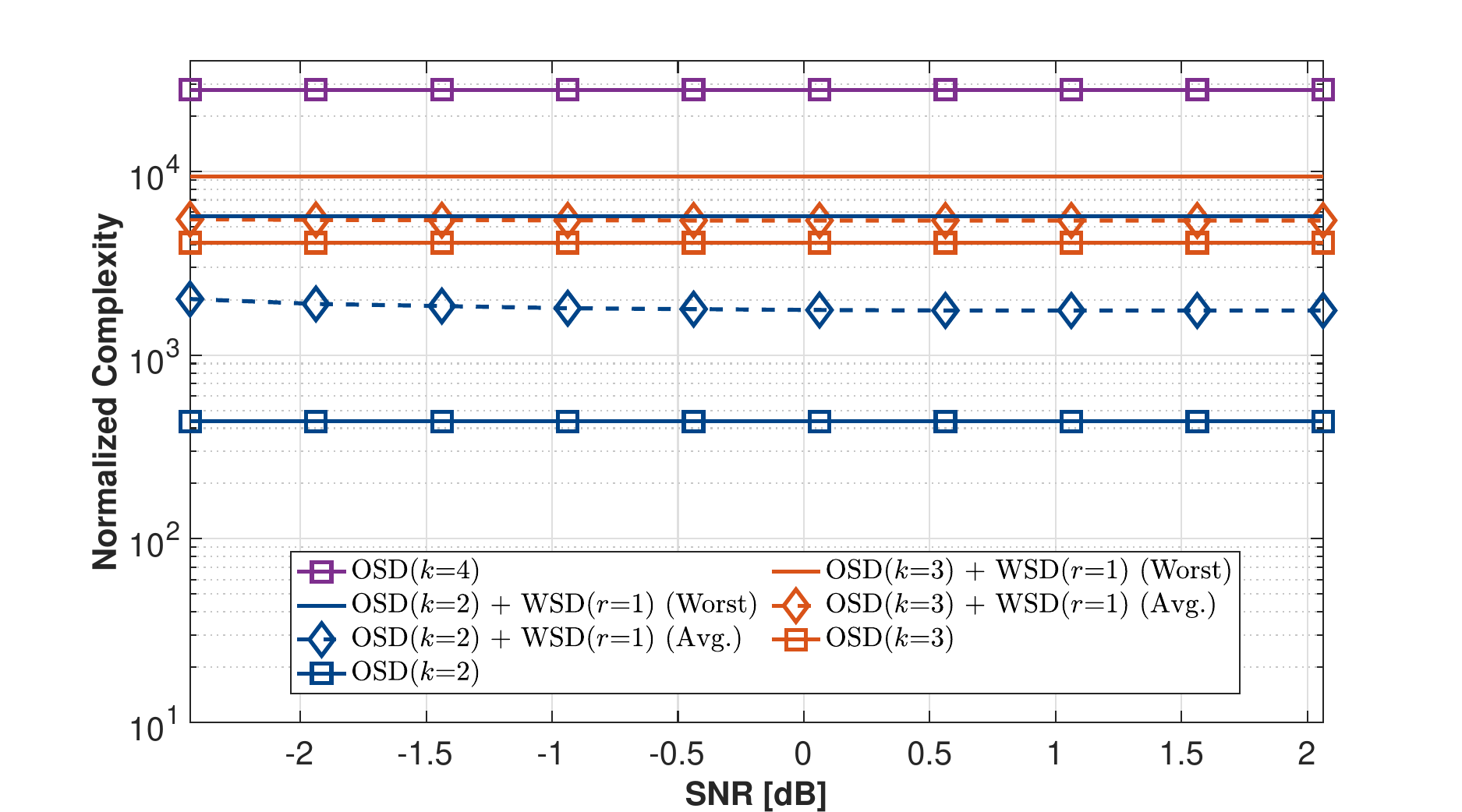}
  \vspace{-3mm}
  \caption{\rev{Average complexity comparison for the $\mathcal{RM}$(128, 29) code under OSD and OSD+WSD methods.}}
  \label{fig:sim_osd_rm_wsd_comp}
\end{figure}

\rev{From the complexity perspective, Table \ref{tab:complexity} summarizes the normalized theoretical complexities, while Figs. \ref{fig:sim_scl_wsd_comp}, \ref{fig:sim_scl_bpc_wsd_comp}, and \ref{fig:sim_osd_rm_wsd_comp} present the corresponding experimental average complexities. The worst-case complexity curves indicate the theoretical upper bound derived in Table \ref{tab:complexity}, calculated assuming $P_{\sf act}=1$ $J=4$.}

As evident in Figs. \ref{fig:sim_scl_wsd_comp} and \ref{fig:sim_scl_bpc_wsd_comp}, the average complexity of SCL+WSD and SCL-BPC+WSD dynamically adapts to the channel conditions, converging to that of the low-complexity initial stage in the high-SNR regime ($P_{\sf act}=P_{\sf e,crc} \approx 0$). Even for $\mathcal{RM}$ code simulation (Fig. \ref{fig:sim_osd_rm_wsd_comp}), where WSD is always activated due to the absence of CRC, \rev{OSD($k=2$)+WSD($r=1$) maintains a complexity orders of magnitude lower than the higher-order OSD($k=4$). This demonstrates that WSD offers a superior trade-off compared to simply increasing the list size of OSD order.}

\section{Conclusion}
In this letter, we presented a two-stage near-ML decoding method that effectively addresses the stringent reliability and latency challenges in finite blocklength regimes for URLLC applications. Our approach adaptively combines a low-complexity initial decoder with an innovative WSD technique.

By exploiting the inherent lattice structure of linear codes and implementing an iterative refinement strategy, WSD achieves near-ML performance with substantially reduced complexity compared to exhaustive methods. The technique's universal applicability to any linear code enhances its versatility. Simulation results confirm robust performance across various code rates, establishing WSD as a promising architectural foundation for future near-ML decoders in reliability-critical systems.

\bibliographystyle{IEEEtran}
\bibliography{letter}

\newpage

\vfill

\end{document}